\documentclass[sigconf]{acmart}

\AtBeginDocument{%
  }

\acmPrice{15.00}

\copyrightyear{2023}
\acmYear{2023}
\setcopyright{rightsretained}
\acmConference[ICBRA 2023]{2023 the 10th International Conference on Bioinformatics Research and Applications (ICBRA)}{September 22--24, 2023}{Barcelona, Spain}
\acmBooktitle{2023 the 10th International Conference on Bioinformatics Research and Applications (ICBRA) (ICBRA 2023), September 22--24, 2023, Barcelona, Spain}\acmDOI{10.1145/3632047.3632072}
\acmISBN{979-8-4007-0815-2/23/09}

\usepackage{siunitx}
\newcommand{\bW}{\boldsymbol{W}}

\newcommand{\bc}{\boldsymbol{c}}
\newcommand{\bo}{\boldsymbol{o}}
\newcommand{\bx}{\boldsymbol{x}}
\newcommand{\bWo}{\boldsymbol{W}^\mathrm{o}} 
\def\tsc#1{\csdef{#1}{\textsc{\lowercase{#1}}\xspace}}
\tsc{WGM}
\tsc{QE}

\begin{document}

\title{Resource-Limited Automated Ki67 Index Estimation in Breast Cancer}

\author{Jessica Gliozzo}
\email{jessica.gliozzo@unimi.it}
\orcid{0000-0001-7629-8112}
\affiliation{%
  \institution{Dipartimento di Informatica, Universit\`a degli Studi di Milano}
  \streetaddress{Via Celoria 18}
  \city{Milan}
  \country{Italy}
  \postcode{20133}
}
\affiliation{%
  \institution{European Commission, Joint Research Centre (JRC)}
  \city{Ispra}
  \country{Italy}
}

\author{Giosuè Marinò}
\email{giosue.marino@studenti.unimi.it}
\affiliation{%
  \institution{Dipartimento di Informatica, Universit\`a degli Studi di Milano}
  \streetaddress{Via Celoria 18}
  \city{Milan}
  \country{Italy}
  \postcode{20133}
}

\author{Arturo Bonometti}
\email{arturo.bonometti@hunimed.eu}
\orcid{0000-0003-3258-0727}
\affiliation{%
  \institution{Department of Biomedical Sciences, Humanitas University}
  \streetaddress{Via Rita Levi Montalcini 4}
  \city{Pieve Emanuele - Milan}
  \country{Italy}
  \postcode{20072}
}
\affiliation{%
  \institution{Department of Pathology, IRCCS Humanitas Clinical and Research Hospital}
  \streetaddress{Via Manzoni 56}
  \city{Rozzano – Milan}
  \country{Italy}
  \postcode{20089}
}

\author{Marco Frasca}
\authornote{Corresponding author}
\email{marco.frasca@unimi.it}
\orcid{0000-0002-4170-0922}
\affiliation{%
  \institution{Dipartimento di Informatica, Universit\`a degli Studi di Milano}
  \streetaddress{Via Celoria 18}
  \city{Milan}
  \country{Italy}
  \postcode{20133}
}
\affiliation{%
  \institution{CINI National Laboratory in Artificial Intelligence and Intelligent Systems}
  \city{Rome}
  \country{Italy}
}

\author{Dario Malchiodi}
\email{dario.malchiodi@unimi.it}
\orcid{0000-0002-7574-697X}
\affiliation{%
  \institution{Dipartimento di Informatica \& DSRC, Universit\`a degli Studi di Milano}
  \streetaddress{Via Celoria 18}
  \city{Milan}
  \country{Italy}
  \postcode{20133}
}
\affiliation{%
  \institution{CINI National Laboratory in Artificial Intelligence and Intelligent Systems}
  \city{Rome}
  \country{Italy}
}

\renewcommand{\shortauthors}{Gliozzo et al.}

\begin{abstract}
The prediction of tumor progression and chemotherapy response has been recently tackled exploiting Tumor Infiltrating Lymphocytes (TILs) and the nuclear protein Ki67 as prognostic factors.
Recently, deep neural networks (DNNs) have been shown to achieve top results in estimating Ki67 expression and simultaneous determination of intratumoral TILs score in breast cancer cells. However, 
in the last ten years the extraordinary progress induced by deep models proliferated
at least as much as their resource demand.
The exorbitant computational costs required to query (and in some cases also to store) a deep model represent a strong limitation in resource-limited contexts, like that of IoT-based applications to support healthcare personnel. 
To this end, we propose a resource consumption-aware DNN for the  effective estimate of the percentage of Ki67-positive cells in breast cancer screenings. Our approach reduced up to $75\%$ and $89\%$ the usage of memory and disk space respectively, up to $1.5\times$ the energy consumption, and preserved or improved the overall accuracy of a benchmark state-of-the-art solution. Encouraged by such positive results, we developed and structured the adopted framework so as to allow its general purpose usage, along with a public software repository to support its usage.
\end{abstract}




\maketitle
\section{Introduction}\label{sec:intro}
The nuclear protein Ki67 has been introduced as a proliferative marker to be used along with Tumor Infiltrating Lymphocytes (TILs) as a feature effectively driving the prediction of tumor progression and chemotherapy response~\cite{taneja2010classical, mao2016TILprognostic}.
Ki67 estimation in surgical pathology is a specific example of a task in which the result of the evaluation bears a significant clinical consequence in many cancer types. Indeed, the pathology practice considers different positivity cut-offs to discriminate between lesions with different overall prognosis or therapy response~\cite{tao2017ki, pena2009does}. For this reason, an high accuracy in Ki67 estimation is essential to avoid over or under-grading of the scrutinized sample. In breast pathology this task may occupy a major role in the  routine of pathologists given the high number of surgical samples produced daily in both referral and peripheral pathology centres, as a consequence of the effectiveness of the woman health screening policies in different countries. These tasks require the segmentation and visual count of cells by pathologists, which is a relatively time-consuming task subject to high intra- and inter-operator variability, related to the experience of each professional~\cite{chung2016interobserver, gomes2014inter}. Furthermore, in the evaluation of all these features it should also be considered the presence of a different degree of intra-tumoral heterogeneity, depending on the cancer histotype and the type of tissue sample~\cite{zilenaite2020independent}. Thus, the development of automatic cell count approaches enables a faster and more reliable diagnosis/prognosis. Indeed, even if the evaluation of a single Ki67 slide may be a relatively fast operation, the in-series estimation of a large number of cancer samples is time consuming and may benefit of automated tools able to
provide and integrate such data in the histopathology report, independently of the computational resources available.
This latter aspect is also of critical importance, as it should be considered that even though screening programs are commonly employed at least in western countries, not every facility may dispose of enough computational
resources to run Deep Learning (DL) software~\cite{harbeck2019breast, park2013history}.
Along with the problem of Ki67 estimation, also the identification of TIL-score is relevant in breast cancer because it has a prognostic role, as a component of the immune system fighting tumor progression~\cite{stanton2016clinical}. TIL estimation from histological images present challenges similar to 
those related to Ki67 (i.e. the task easily becomes time-consuming and operator-dependent), thus also this activity can take advantage from the application of DL.


The successful results of deep convolutional  models are not surprising, since these models proved to be one of the most effective and flexible tools to tackle healthcare problems~\cite{bengio2017deep, egger2022medical}. These methods are currently applied to analyse a wide spectrum of
medical data types, such as
(I) clinical images,
(II) biosignals,
(III) high-dimensional 
omics data, and
(IV) Electronic Health Records,
achieving state-of-the-art performance on 
several tasks, including the recognition of mitoses, nodal metastases, or the measurement of prognostic markers~\cite{Hanna2021, klein2021machine, Xu2021, Lee2021}.

The application of DL models is also pervasive in the analysis of breast cancer histopathological images, where it is effective to tackle different tasks: image classification~\cite{debelee2020survey, aljuaid2022computer, meirelles2022effective},  cell contours and nuclei detection~\cite{mridha2021comprehensive} and Ki67-index estimation.
Regarding breast cancer image classification, state-of-the-art methods are based on convolutional neural networks. Nawaz et al.~\cite{nawaz2018multi} proposed the use of the convolutional architecture DenseNet to perform multi-class classification of histopathological images from the BreakHis dataset~\cite{spanhol2015dataset}, reaching an accuracy of 95.4\%.
Xie et al.~\cite{xie2019deep} used the Inception\_V3 and Inception\_ResNet\_V2
networks to
classify
breast cancer histopathological images through transfer learning,
outperforming
many previously proposed architectures (AlexNet, CSDCNN, LeNet).
Zhu et al.~\cite{zhu2017extended} used a fully-CNN exploiting convolution and deconvolution layers to directly regress a density map showing the position of cells. Interestingly, they evaluated the model also on histopathological breast lymph node images. Xue et al.~\cite{xue2016cell} use AlexNet and ResNet with Euclidean loss to predict the number of cells in image patches.
The use of DL for cell counting is so well-established in literature that an ImageJ plugin~\cite{falk2019u} now enables the use of U-Net models to perform cell detection on generic biomedical image data.
Finally, the estimation of Ki67-score from immunohistochemical images is another task where DL is largely adopted. To this end, Saha et al.~\cite{saha2017advanced} initially proposed to use a Gamma mixture model with Expectation-Maximization to identify seed points and patch selection, which are fed to a CNN model having a custom decision layer based on decision trees. The method achieved 91\% of F-score.
Zhang et al.~\cite{zhang2018tumor} explored the use of CNN for Ki67 image classification, where a probability heatmap can be obtained from repositioning the classified patches to their original location. The ratio of tumor cells in the heatmap can be used as an indicator for Ki67 expression. Moreover, they explored the use of single shot multibox detectors to detect Ki67 positive and negative cells. Fulawka et al.~\cite{fulawka2022assessment} applied a CNN DenseNet model with fuzzy interpretation to obtain a binary mask, which is then used to segment breast cancer cells and compute Ki67 index. Negahbani et al.~\cite{negahbani2021pathonet} introduced a pipeline for cell classification and detection of both Ki67 and TILs that exploits a U-Net architecture with novel residual dilated inception modules (see section~\ref{sec:algo} for further details).
However, the resources required to run DL applications are not trivial, being these models typically overparameterized~\cite{Allen19}, and therefore requiring a high amount of computational resources. We refer here to the size of the learned DL models, and not to the ever-growing size of the datasets to be processed~\cite{li2021systematic}, {or the size of the single images, as these issues have been addressed in literature (see, e.g.,~\cite{hirahara2021effect} and the use of tiles/patches~\cite{Berman2021pathML})}. Just to state some examples, depending on their specific implementation, the space needed to store a CNN trained for image classification purposes, such as the previously mentioned DenseNet, ResNet, or AlexNet models, varies from tens to hundreds of megabytes; this requirement jumps up to several tens gigabytes if we consider modern Large Language Models (e.g., 11 billions of parameters for some variants of the T5 Text-To-Text Transformer Model~\cite{T5_20}, meaning around $44$ terabytes of RAM when using $4$ bytes per parameter). To face the issue related to the excessive resource usage, recent works focused on specifically designing deep models for mobile devices~\cite{Zhang18}, which however, despite their effectiveness, is applicable to novel architectures but not to the existing ones. Indeed, most available DNNs have been pre-trained (often using a more than remarkable amount of resources) with the aim of deploying them onto standard computing facilities.
Such aspect can easily hamper the adoption of DL-based medical software, especially on devices with limited resources, like smartphones or IoT-devices~\cite{Garca19,Liu15,Abdellatif20}. 
Further, this is even more relevant in developing countries, characterized by poor economic conditions---mainly in the rural areas---and often even by shortage of physicians, which would highly benefit from AI systems to support clinical decisions. Low-resource DL models can improve the exploitation and applicability of the basic wearable networked devices available, and foster a speed-up toward revolutionary changes in the healthcare systems of these countries~\cite{GUO18}. {On the other side, cloud-based solutions might be used for hosting large DL models in order to query them online. However, apart privacy and security issues induced by sensible data, it would also require an active Internet connection, which might be a problem in poor rural areas. Moreover, Internet and cloud-based solutions have a cost which has to be taken into account, in relation with the (scarce) available budget. }

In such a context, this study proposes an automated AI method for cell classification and detection of Ki67 and TILs, with a particular attention to the resource usage. We show our solution to be competitive with the state-of-the-art solution,  PathoNet~\cite{negahbani2021pathonet}, a CNN shown to be top-performing in classifying breast cancer images belonging to three classes \emph{Ki67- immunopositive}, \emph{Ki67- immunonegative} and \emph{tumor infiltrating lymphocytes}, while reducing its RAM and disk requirements up to $4\times$ and $9\times$, respectively{, along with its energy consumption}.  
Moreover, once verified the effectiveness of our methodology, we have done a further step towards the realization of a more general framework to reduce the resource demand of existing pre-trained DL models. {In this sense, PathoNet can be considered as a use case of such a methodology, relatively `toy' for its memory size of around {$13 MB$}, but we will argument in the discussion at the end of the paper how recent researches already proposed DNNs for the same problem requiring several hundreds of megabytes of RAM. 
Summarizing, we provide} a Python module to 1) first let the existing pre-trained model to undergo lossy compression of its layers, and then lossless storage of the compressed layers; 2) to run in main memory the model in the resource-saver format and efficiently querying it directly in such a format, without need to reconvert it to the original uncompressed format.

The paper is organized as follow: 
Sect.~\ref{sec:methods} illustrates the used data, model and overall methodology. The obtained results are described in Sect.~\ref{sec:results} and discussed in Sect.~\ref{sec:discussion}. Some concluding remarks end the paper.

\section{Methods}\label{sec:methods}
This section depicts the application of the proposed framework to the considered case study. More precisely, Sect.~\ref{sec:data} describes the used dataset, while Sect.~\ref{sec:algo} illustrates the framework adopted to learn the low-resources model, as well as how to run queries in the compressed storage format.
\subsection{Problem Definition and Data}
\label{sec:data}
We requested to the authors the breast cancer invasive ductal carcinoma dataset (SHIDC-B-Ki-67-V1.0)\footnote{\url{https://github.com/SHIDCenter/PathoNet}}, which is already divided
in a train and a test set (containing $1656$ and $700$ images, respectively).
Each image is associated with a JSON file containing the nuclei coordinates (position on X and Y axes of the cell center) and the ground truth label for each nucleus (where 1=\emph{Ki67-immunopositive}, 2=\emph{Ki67-immunonegative} and 3=\emph{TIL}).
From the provided training set, we created a validation set randomly selecting $20 \%$ of the
labeled images.
We checked that the obtained train (composed by the remaining $80 \%$ of the images) and validation set had an average number of cells per image (i.e. ``avg./IMG'') similar to the values published in Table 2 of the reference paper~\cite{negahbani2021pathonet}. The obtained values of avg./IMG are presented in Table~\ref{tab:stats_cells}. The train set undergoes data augmentation by flipping images w.r.t.\ X and Y axes and applying rotations, as done in the reference paper.

\begin{table*}[ht]
\centering
\small
\caption{Average number of cells per image (avg./IMG) and number of annotated cells (\# cells) for each set, presented across cell labels, where immunopositive (Ki67 +) and immunonegative (Ki67 -) represent the result of Ki67 staining.}
\label{tab:stats_cells}
\begin{tabular}{lS[table-format=2.2]r|S[table-format=2.2]r|S[table-format=2.2]r|S[table-format=2.2]r}
\toprule
 & \multicolumn{2}{c|}{\textbf{\begin{tabular}{@{}c@{}}Training+ \\ Validation \end{tabular}}} & \multicolumn{2}{c|}{\textbf{Training}} & \multicolumn{2}{c|}{\textbf{Validation}} & \multicolumn{2}{c}{\textbf{Test}} \\
 
 & \multicolumn{1}{c}{avg.\slash IMG} & \multicolumn{1}{c|}{\# cells} & \multicolumn{1}{c}{avg.\slash IMG} & \multicolumn{1}{c|}{\# cells} & \multicolumn{1}{c}{avg.\slash IMG} & \multicolumn{1}{c|}{\# cells} & \multicolumn{1}{c}{avg.\slash IMG} & \multicolumn{1}{c}{\# cells} \\
 \midrule
\textbf{Ki67 +} & 21.21 & 35106 & 20.88 & 27664 & 22.48 & 7442 & 22.51 & 15755 \\
\textbf{Ki67 -} & 45.31 & 75010 & 45.04 & 59683 & 46.31 & 15327 & 46.63 & 32643 \\
\textbf{TIL} & 1.88 & 3112 & 1.81 & 2402 & 2.15 & 710 & 1.97 & 1380 \\
\bottomrule
\end{tabular}
\end{table*}

\subsection{Algorithm}\label{sec:algo}
\subsubsection{Notation and preliminary definitions}
Most memory and disk requirements of a convolutional neural network (CNN) are due to the storage of weight connections for each layer in the model. Thus, the compression of an existing CNN mainly
maps to the problem of finding out
a succinct  approximation $\bW$ of a given matrix/tensor $\bWo$ representing the learned connection weights of a network layer. The \textit{compression ratio} is defined as the ratio $\psi$ between
the sizes of the uncompressed and compressed matrix,
that is $\psi = \mathrm{size}(\bWo)/\mathrm{size}(\bW)$, where $\mathrm{size}(x)$ is the memory size of $x$.
In general, uppercase boldface symbols will denote matrices, and the corresponding lowercase letters will refer to matrix entries (e.g., $w^\mathrm{o}$ will be an entry of $\bWo$). Vectors---precisely, row vectors---will be rendered using italic boldface (e.g., $\bx$).

\subsubsection{Compression Framework}
{The first phase of the proposed methodology concerns the reuse of an existing pre-trained model, as common nowadays given} the large number of publicly available pre-trained models developed for various applications. To show the power and flexibility of the proposed approach  we thereby compress the pre-trained PathoNet network, retrieved from the same repository holding the previously mentioned data.

\paragraph{The PathoNet model}
PathoNet is a CNN proposed for the accurate detection and count of tumoral cells, appropriately stained for Ki67 and TILs, from biopsy images of malignant breast tumors. The cells present in these images have been labelled as Ki67 positive tumor cells, Ki67 negative tumor cells, and lymphocytes infiltrating the cancer area. The model is designed to detect and classify cells according to these three classes~\cite{negahbani2021pathonet}.

To deal with the size variability of cells from image to image, PathoNet is mostly composed of \textit{dilated}
inception layers. On the one hand, this
solves the problem of choosing a fixed kernel size by using different kernel sizes in one module;
on the other one, it enlarges the network structure
without suffering from the vanishing gradient problem when increasing the number of kernels. In particular,
residual dilated inception modules (RDIMs) are used
in the encoder and decoder part of the network.
Each of these modules consists of
two parallel parts, the first composed by two stacked convolutional layers with kernel size $3 \time 3$, and the second built by stacking two $3 \times 3$ dilated convolutional layers with dilation 4. To reduce the number of parameters, and consequently the possibility of overfitting, the outputs of these two parts are not concatenated but summed up. Overall, PathoNet utilizes a U-Net-like structure~\cite{UNET}, where most convolutional layers are replaced by RDIMs.
The structure of the network is the following: the input layer is initially processed using two convolutional layers, in turn stacked over four encoder RDIMs and four decoder RDIMs; the latter are followed by three $1 \times 1$ convolutional layers with linear activation function, used to produce the three-channel output of the model (see~\cite{negahbani2021pathonet}, Figure 5, for a visual representation).
\paragraph{Lossy compression of network layers}
This step is crucial to achieve good compression ratios while not affecting the model accuracy. Among the vast plethora of compression techniques proposed in literature (see, e.g.~\cite{Deng20} for a survey), we designed and exploited the most suitable techniques for the structure of PathoNet.
We did not consider weight or structural pruning, since the model network structure is already tuned by the authors, with a well-conceived and precise organization of the individual blocks. Weight pruning, for instance, is rewarded when attaining high sparsity levels ($> 0.5$), which in turn allow the usage of compressed formats such as CSC. However, an
excessive pruning of convolutional layers can highly deteriorate the predictive capabilities of the model~\cite{MARINO23}. 
Additionally, our aim is also to not slower the model execution, along with obtaining its space reduction. Accordingly, we operated a weight quantization of convolutional layers, consisting
in building the matrix $\bW$ using
a limited number of distinct weights, each represented using less bits than each entry in $\bWo$.
The idea is to cast connection weights into categories and substituting all weights in a category with a representative. This approach, named weight sharing (WS), allows to save room when combined with the Index Map storage format, which consists in  replacing weights with their category/representative index, at the cost of storing separately the vector of representative weights. The advantage of this format is that it only adds one level memory access and almost preserves the dot time efficiency of the original model. It is worth pointing out that in case the model to compress contains also other types of layers, e.g., fully-connected layers, specific and more suitable lossy compression and lossless storage approaches can be leveraged,
exploiting for instance
pruning+quantization compression and address map storage~\cite{HAMICPR,RRPR21}.

The strategy used to group weights and to select representatives distinguishes the four state-of-the-art quantization algorithms considered in this study and described in the following.
\begin{list}{}{}
\item \textit{Clustering-based WS} (CWS).
This strategy groups weights into $k$
clusters via the $k$-means algorithm~\cite{mcqueen},
and uses the resulting centroids $\{c_1, \dots, c_k \}$ as representatives to replace the weights in the corresponding cluster~\cite{Han15}.
\item \textit{Probabilistic WS} (PWS). This technique is based on the \textit{Probabilistic Quantization} method~\cite{HAMICPR}, in which a randomized algorithm transforms each weight $w^\mathrm{o} \in \bWo$ in one of $k$ distinct representatives $c_1, \ldots, c_k$. A nice feature of this method consists in the fact that the obtained $\bW$ can be seen as the value of an unbiased estimator for $\bWo$ (see~\cite{HAMICPR} for details).
\item \textit{Uniform Quantization} (UQ).
In this scheme, which achieved top compression performance in recent applications to CNNs compression~\cite{Choi20}, representatives are selected by uniformly partitioning the entire weight domain into $k$ subintervals\footnote{Note that the actual number of subintervals $k$ can be lower than the input value due to the internal selection of the $\delta$ hyperparameter of the method (see~\cite{Choi20} for further details).}.
Such a selection has been proven to yield an output entropy which is asymptotically smaller than that of any other quantizer, regardless of the source statistics, when the source density function is sufficiently smoothed~\cite{GishH1968}.

\item \textit{Entropy Constrained
Scalar Quantization} (ECSQ): This is a technique leveraging an iterative optimization algorithm to determine the optimal number of groups. It is driven by the joint optimization of the expected value for the quantization distortion (a measure of the distance between $\bWo$ and $\bW$) and the entropy of the resulting discrete distribution for representative weights~\cite{chou1989entropy}. 
\end{list}

A retraining phase is finally applied after quantization, ensuring that the updated weights always assume values in the set of representatives~\cite{Han15}.
\paragraph{Lossless storage of the compressed network}
The third phase of our framework is the design of a suitable room-saver format for the model, tailored for the compression schemes used in the previous step, and able to perform the model execution without re-expanding the network. As mentioned above, 
an efficient solution to exploit the quantized tensors is represented by the Index Map (IM) format~\cite{Han15}. Representatives are stored in a vector $\bc =\{c_1, \ldots, c_k\}$, whose indices
are entries of a new matrix/tensor $\boldsymbol{M}$.
Thus, if $w^\mathrm{o}\in \bWo$ is associated with centroid, say, $c_2$, then the corresponding entry in $\boldsymbol{M}$ is set to $2$.
When $\bWo$ (and accordingly $\boldsymbol{M}$) has dimension $n\times  m$, denoted by $b$ and $\bar b$ the number of bits used to store one entry of $\bWo$ and of $\boldsymbol{M}$, respectively, the \textit{compression ratio} obtained is: 

\begin{displaymath}
\psi = \frac{bnm}{\bar b nm + bk}
\end{displaymath}

For instance, when $k\leq 256$, $\bar b=8$ is enough to represent $2^8=256$ different indices, and assuming a typical FP32 format for $\bW^\mathrm{o}$ ($b=32$), the compression ratio would be $\psi \approx 4$. We remark that this format does not induce any information loss, while needing only one additional memory access to retrieve a given weight.
\paragraph{Model inference in the resource-saver format}
The final step of our framework is the computation of matrix/tensor products directly in the compressed format used at previous step. Without loss of generality, we can assume that the layer weights are represented by a matrix. To compute the output $\bo = \bx \cdot \bW$ of a given compressed layer with weight matrix $\bW$ on the input $\bx$, that is $o_i = \sum_j x_j w_{ji}$ using IM format, we perform $o_i = \sum_j x_j c_{m_{ji}}$. From a memory consumption standpoint, this does not need to expand the compressed matrix, and it still keeps the model memory footprint $\psi$ times smaller than the original one. 
\paragraph{Implementation}
The PathoNet source code was implemented in Python 3, using Tensorflow and Keras. Our compression techniques and the retraining procedures have been implemented in the same programming environment. Our implementation, available on GitHub\footnote{\url{https://github.com/GliozzoJ/pathonet_compression}}, allows: (i) to perform the compression and retraining of PathoNet with different quantization techniques; (ii) to compute the compression ratio (relative to the memory space); (iii) to evaluate the compressed model space on disk and the prediction time ratio compared to the uncompressed version of PathoNet. The repository also contains a Jupyter notebook allowing the replication of our results via direct execution of the compressed models relying on IM representation and another notebook to estimate energy consumption.

\section{Results}\label{sec:results}

We tested the
four considered quantization approaches (i.e., CWS, PWS, UQ, and ECSQ; see Sect.~\ref{sec:algo})
on the breast cancer dataset provided by the authors of PathoNet. As mentioned in Sect.~\ref{sec:data}, we executed an holdout procedure in which the training set is exploited to retrain the model after quantization and a validation set is used to tune a set of hyperparameters by means of grid search.
Indeed, the approaches adopted to compress the network present some hyperparameters that influence the obtained performance.
One of them is the number $k$ of groups, which has a direct effect on the final size of the compressed model. 
Moreover, since we have to retrain the network after the quantization step, all the classical hyperparameters related to the training of a neural network (e.g., learning rate, batch size, patience, etc.) play an important role. In particular, we have 
considered the cumulative learning rate $clr$ and the number of groups $k$, since they impacted more on the model accuracy.
Once the best model is selected (see Sect.~\ref{sec:data}), the corresponding hyperparameters are used to train the network on the train and validation sets, and the generalization performance of the compressed model is assessed on the test set.

\subsection{Evaluation Metrics}
The successful application of a compression strategy should lead to a model that
(I) retains similar generalization performance w.r.t. the uncompressed model,
(II) leads to a reduction in terms of space occupancy, and
(III) keeps a reasonable execution time when used to make inferences on the test set.

In particular, the generalization performances are evaluated in terms of 
F1-score, RMSE (Root Mean Squared Error) and aggregated cut-off accuracy for TILs and Ki67 (as defined in the reference paper). 
Moreover, two additional metrics are used:
\begin{itemize}
    \item \textit{Compression ratio:} the ratio of the memory size needed by the uncompressed over the compressed model (cfr.\ Sect.~\ref{sec:algo});
    \item \textit{Time ratio:} the ratio between evaluation times on the test set of the uncompressed over the compressed model.
\end{itemize}

\subsection{Hyperparameters optimization}
The train and validation sets (Section~\ref{sec:data}) are used to perform the tuning of hyperparameters, i.e., the cumulative learning rate $clr$ for the fine tuning of weights after quantization (Section~\ref{sec:algo}), and number of clusters $k$, by means of grid search. Obviously, the validation set is not augmented in this phase. The best combination of hyperparameters is the one that gives the lowest RMSE on the validation set. Then the best model is retrained on the complete augmented training set.

As outlined in~\cite{HAMICPR}, the cumulative learning rate needs to be smaller than the learning rate used to train the original model: accordingly, the grid for $clr$ has been set to $[0.001, 0.0001, 0.00001, \linebreak 0.000001]$.  
The number of groups $k$ has been chosen in $[256, 1024, 4096]$ for all compression methods, with the first choice ensuring using 1 byte for each index, and the other 2 choices ensuring lower compression but potentially higher performance. Note that this bidimensional grid yields $12$ combinations for each method, for a total of $48$ experiments; this prevented us from using more refined grids. 
After choosing the best couple, a final retraining using all the augmented training set produces the compressed model. The cumulative learning rate and the number of groups selected at the end of tuning are showed in Table~\ref{tab:tuning_params}. The network configuration was kept as in the original PathoNet model, whenever possible, and all the other experimental details are reported in our public repository\footnote{\url{https://github.com/GliozzoJ/pathonet_compression}}. 

\begin{table}[t]
\centering
\caption{Values of cumulative learning rate (\textit{clr}) and number of groups $k$ selected by the tuning process. 
}
\label{tab:tuning_params}

\begin{tabular}{lllll}
\toprule
Quantization & CWS & PWS & ECSQ & UQ \\
\midrule
\textit{clr} & 0.00001 & 0.0001 & 0.00001 & 0.0001 \\
\textit{k}   & 1024    & 4096   & 4096    & 4096 \\
\bottomrule
\end{tabular}

\end{table}


\subsection{Experimental results.}
The generalization performance, in terms of F1-score for the three classes, RMSE and aggregated cut-off accuracy for Ki67-index and TIL-score is presented in Table~\ref{tab:exp_perf} and compared to the same metrics computed on the original uncompressed PathoNet model~\footnote{The results reported for the uncompressed network in Table~\ref{tab:exp_perf} are different from the ones showed in the reference paper~\cite{negahbani2021pathonet}. We contacted the authors of PathoNet and they agreed with the correctness of the F1-score results using dataset SHIDC-B-Ki-67-V1.0. Moreover, we implemented the function to compute the RMSE and cut-off accuracy, which is now part of the PathoNet package available on GitHub (\url{https://github.com/SHIDCenter/PathoNet/blob/master/evaluation.py}).}. 

\begin{table*}[ht]
\centering
\small
\caption{Generalization performance of the uncompressed (first row) and compressed PathoNet models. Ki67+ stands for immunopositives, Ki67- for immunonegatives, while the last two columns respectively represent the RAM compression  and the evaluation time ratios (the higher, the better). The best results for each metric are highlighted in bold. 
}
\label{tab:exp_perf}
\begin{tabular}{l|ccc|cc|cc|cc}
\toprule
& \multicolumn{3}{c|}{\textbf{F1-score}} & \multicolumn{2}{c|}{\textbf{RMSE}} &\multicolumn{2}{c|}{\textbf{Cut-off Accuracy}} \\
\textbf{Experiment} & \textbf{Ki67+} & \textbf{Ki67-} & \textbf{TIL} & \textbf{Ki67-index} & \textbf{TIL-score} & \textbf{Ki67-index} & \textbf{TIL-score} & \textbf{Space} & \textbf{Time}\\ 
\midrule
Uncompressed & \textbf{0.853} & 0.776 & 0.348 & 0.050 & 0.054 & 0.913 & 0.826 & - & - \\

CWS & 0.852 & 0.774 & 0.355 & 0.054 & 0.044	& 0.913	& 0.870 &1.942 & 0.833 \\ 

PWS & 0.848	& 0.774	& 0.351 & 0.053	& \textbf{0.021} & 0.913 & \textbf{0.957} &1.789 & 0.838 \\ 

ECSQ & 0.852 & 0.776 & 0.375 & 0.054 & 0.039 & 0.913 & 0.913 &1.789 & 0.837 \\

UQ & 0.852 & 0.775 & \textbf{0.378} & 0.056 & 0.036 & 0.913 & 0.913 & 1.916 & 0.835 \\

CWS - k=256 & \textbf{0.853} & 0.778 & 0.365 & 0.053 & 0.041	& 0.913	& 0.870 &\textbf{3.937} &\textbf{0.843} \\

PWS - k=256 & 0.835 & 0.746 & 0.193 & 0.050 & 0.028	& \textbf{0.957} & \textbf{0.957} &\textbf{3.937} & 0.839 \\

ECSQ - k=256 & 0.848 & \textbf{0.781} & 0.355 & \textbf{0.049} & 0.023 & 0.913 & 0.913 &\textbf{3.937} & 0.837 \\

UQ - k=256 & 0.852 & 0.777 & 0.374 & 0.055 & 0.038 & 0.913 & 0.913 &\textbf{3.937} & 0.840 \\

\bottomrule
\end{tabular}
\end{table*}

As we can see from the results of this first set of experiments (rows 2--5 in Table~\ref{tab:exp_perf}), the compressed networks achieve comparable performance w.r.t.\ the original PathoNet model in terms of F1-score for the classes Ki67 immunopositive and immunonegative.  Interestingly, the compression methods UQ and ECSQ obtain a significant improvement in F1-score for the TIL class ($3\%$ for UQ and slightly lower for ECSQ). 
Considering RMSE, the performances of compressed networks are slightly worse for the Ki67-index, but they achieve always better results for the TIL-score. 
Moreover, all the compressed networks match the uncompressed ones for the Ki67-index cut-off accuracy while always consistently improving the corresponding metric for TIL-score. In particular, the improvement in cut-off accuracy for TIL-score ranges from $4.4\%$ to $13.1\%$, depending on the applied quantization approach.  
Overall, all the compressed models almost halve the space occupancy in RAM while bringing a slow down during execution of less than $20\%$.

\paragraph{Best compression}
As shown in Table~\ref{tab:tuning_params}, the grid search process led to the selection of an high number of groups $k$, which was equal to 4096 in most cases. The performance of the compressed models is competitive with the original PathoNet model, and some metrics are often better (especially the ones related to TILs). This behaviour is expected, since an higher number of representatives gives better chances to preserve the network structure. On the other hand, it is interesting to evaluate if a lower number of representatives can lead to competitive results while significantly reducing the space occupancy in main memory.
From a practical standpoint, a user could have only limited computational resources available and willing to still use a compressed CNN even at the expenses of a marginal decrease in generalization performance.
To test this situation, we executed again the experiments performed in the previous section but avoiding the model selection process. In particular, the number of groups is fixed to $256$ for all quantization methods and the
cumulative learning rate
as the best value selected by means of grid search in the previous set of experiments (see Table~\ref{tab:tuning_params}). The other hyperparameters remained unchanged.
The results are showed in Table~\ref{tab:exp_perf} (rows 6--9). Quite surprisingly, compressed PathoNet models in this setting tend to preserve or even improve their performance when using more representatives (see, e.g., the CWS method), which has the appreciated benefit that the space compression is still increased, namely to $\approx 4 \times$ the original uncompressed PathoNet model.
An exception is the PWS method, which shows a lower ability to select informative representatives when their number is limited, confirming the results in~\cite{MARINO23}.
On the other side, CWS method exhibited an overall higher stability and effectiveness in choosing the representative weights, as confirmed by the the fact that it performed best during model selection when not using the maximum $k$ allowed (Table~\ref{tab:tuning_params}). 

\section{Discussion}\label{sec:discussion}
\begin{figure}
\centering
\includegraphics[width=0.3\textwidth]{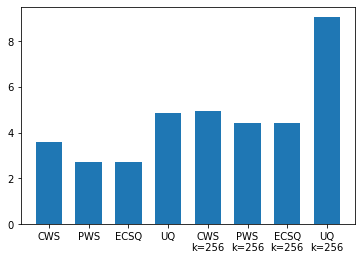}
\caption{Disk compression ratio of the tested methods with respect to the original size of PathoNet (12.802 MBytes).}
\label{fig-compression-ratio}
\Description[Disk compression of tested approaches.]{Uniform Quantization using 256 groups allows to obtain a model 9 times smaller with respect to the original Pathonet model on disk. Clustering-based weight sharing , Probabilistic weight sharing and Entropy Constrained Scalar Quantization using 256 groups lead to a 4 times reduction in disk usage. Finally, the methods considering a tuned value for the number of groups achieve a reduction in disk usage of 4 times for Uniform Quantization and Clustering-based weight sharing, while Probabilistic weight sharing and Entropy Constrained Scalar Quantization obtains a reduction of 2 times.}
\end{figure}

In this research, we focused on the problem of the automatic computation of Ki67 and TIL indices in breast cancer, with the primary goal of at least preserve the top performance obtained for this task in the literature, along with a special attention to limit the resulting model resource usage, to not hamper its practical applicability. 
We showed how to obtain a CNN exhibiting performance competitive with PathoNet (reference CNN for the problem), while yielding a model much more resource-cautious, through a novel compression framework. Our solution is around $4\times$ smaller in terms of memory footprint 
 and $9\times$ in terms of disk size (Figure~\ref{fig-compression-ratio}), with reference to PathoNet, while still performing the same or better. 
This is at first of clinical interest, being Ki67 pivotal in the definition of patients' treatment and for the evaluation of their prognosis~\cite{dowsett2011assessment}. Similarly, the number of TILs show a positive correlation with patients survival and therapy response. In fact at least in certain subtypes of breast cancer, a higher number of TILs indicates a higher activation of tumor-suppressing adaptive immunity and an increased rate of response after adjuvant anthracycline-based chemotherapy~\cite{salgado2015evaluation}.
For example, in lung cancer, the evaluation of PDL-1 immunohistochemical positivity became critical after the demonstration of the efficacy of anti-PDL1 drug Pembrolizumab~\cite{gandhi2018pembrolizumab}. The inclusion of patients in im\-munochemother\-a\-py protocols with Pembrolizumab monotherapy or its combinations passes through PDL1 positive cell count, since inter-observer variability can easily shift the therapeutic plan due to the low amount of positive cells needed to reach the TPS cut-off~\cite{cooper2017intra}.
Indeed, attempts to use DNN for TPS computation in lung cancer already exist in  literature~\cite{wang2021dual}. 
    
Moreover, given the growing knowledge about disease molecular therapy targets and interest in Precision Medicine, a future increase in the number of routinely analyzed immunohistochemical predictive and prognostic markers becomes a reasonable educated guess~\cite{fassan2020current}. In this context, the application of ML-based models on prognostic/predictive immunohistochemical panels will shorten the time required for their evaluation, and increase the overall accuracy of the tests. The employment of such models will likely reduce the costs and time related to the evaluation of each biomarker, therefore mitigating the pathologist workload; reducing the resource demand of such computational  models will thereby still favour their applicability.

{Secondly, apart the relevance of cell count operation emphasized so far, our work induces a second benefit related to the model compression, in terms of both RAM and disk occupancy reduction}. In general, the disk space required to store a CNN is lower than the amount of RAM needed to load and query the model, due to optimized formats available to serialize models on disk (e.g., the disk serialization provided by the lzma Python module\footnote{\url{https://docs.python.org/3/library/lzma.html}}). {The disk storage reduction (up to $9\times$, UQ quantization method),  represents an advantage in situations in which the serialized representation of models} is used to share the latter among several actors in a distributed framework. This happens notably in the \emph{federated learning} setting~\cite{li2020federated}, characterized by a privacy-preserving communication loop in which edge computing devices train ``partial'' machine learning models on locally acquired data and send them to a centralized server that merges them and shares with all devices the resulting ``global'' model. Having the possibility to send compressed models back and forth would lead in this case to a valuable saving of network bandwidth, {which is} the bottleneck resource.

Motivated by the promising results obtained, we have made our compression methodology ready and general enough to be applied in other domains with similar needs. {In this sense, PathoNet can be seen as a `toy example' for stating the usefulness of our methodology, being its memory footprint slightly less than 13 MB, taking into account 4 bytes for each model parameter as a floating point number. Nevertheless, for the computation of Ki67-index current researches have also proposed much bigger models: see e.g.~\cite{fulawka2022assessment}, where an ensemble of three DenseNet121 models is used to provide a final prediction,  which roughly requires 320 MB of RAM. Although in small scale, we give an idea of how much the compression of PathoNet impacts on the energy consumption. We {leveraged} the Python package \textit{codecarbon} and obtained $0.000573$ kWh used for querying the original model on $500$ images}, and $0.000380$ kWh to query the compressed one (i.e. UQ, $k=256$).\footnote{Energy consumption was computed on Linux-6.2.6-76060206-generic-x86\_64-with-glibc2.35, CPU Intel(R) Core(TM) i7-9750HF 2.60GHz, GPU NVIDIA GeForce RTX 2060.}
{Our repository provides a dedicated Jupyter notebook to specifically compare the energy consumption (see Sect.~\ref{sec:methods})}, in addition to the code able to reproduce all the experiments proposed in this study, and to the scripts allowing to automatically load serialized saved models, deserialize them, and run their execution.

\section{Conclusions}
This study introduces an automated DL approach for the Ki67 - immunonegative, Ki67 - immunopositive, and TILs cell detection on stained images, with a specific attention to the resulting model resource demand. Our framework  exhibits performance competitive with state-of-the-art models for the estimation of Ki67 and TIL indices in breast cancer. 
Further, it successfully tackles the problem of reducing model size and computational resource need, to enhance its applicability in low-resources context, and to limit the energy consumption.  With reference to a state-of-the-art top-performing solution, we obtained a model up to $4\times$ and $9\times$ smaller in terms of RAM and disk space respectively,
while reducing the energy consumption of around $1.5\times$, and substantially preserving classification accuracy. {Once favorably demonstrated the effectiveness of our approach in the estimation of Ki67 and TIL scores, we have done a further step to settle a general framework for compressing pre-trained DL models, that  
coupled with a publicly available repository, allows to extend our approach to other problems and deep models}. This study serves thereby as a potential reference for non-expert users who need to downsize existing AI tools based on deep or convolutional neural networks, avoiding building compact DNNs from scratch, mainly for problems where the training has very high costs. Our solution allows to extend existing models to contexts where sufficiently powerful hardware is not available or where devices have inherently few computational resources. 
Our pipeline can handle both fully-connected and convolutional layer in compression step, and it is not limited to a given storage format in the final compressed layer representation step.
The main limitation of the overall framework lies in its applicability to only deep models (which however cover the majority of application domains), and among them, to feed-forward architectures, excluding for instance recurrent neural networks (RNNs), just to state an example. A potential future development of this study would indeed extend the present solution to also support such models.

\begin{acks}
This work was supported by the Italian MUR PRIN project ``Multicriteria data structures and algorithms: from compressed to learned
indexes, and beyond'' (Prot. 2017WR7SHH).
Part of this work was done while D.\ Malchiodi was visiting scientist at Inria Sophia-Antipolis/I3S CNRS Université Côte d’Azur (France).
We thank M.Sc. Farzin Negahbani and his research team at Shiraz University for the prompt assistance with the PathoNet package.
\end{acks}

\bibliographystyle{ACM-Reference-Format}
\bibliography{main}


\end{document}